\documentclass[pdflatex,sn-mathphys]{sn-jnl}

\makeatletter
\patchcmd{\ps@headings}
{\hbox to \hsize{\hfill Springer Nature 2021 \LaTeX\ template\hfill}}
{\hbox to \hsize{}}
{}
{}
\patchcmd{\ps@headings}
{\hbox to \hsize{\hfill Springer Nature 2021 \LaTeX\ template\hfill}}
{\hbox to \hsize{}}
{}
{}
\patchcmd{\ps@titlepage}
{\hbox to \hsize{\hfill Springer Nature 2021 \LaTeX\ template\hfill}}
{\hbox to \hsize{}}
{}
{}
\makeatother

\jyear{2021}%

\theoremstyle{thmstyleone}%
\theoremstyle{thmstyletwo}%

\theoremstyle{thmstylethree}%

\usepackage{graphicx}
\usepackage{anyfontsize}
\graphicspath{ {./images/} }

\begin{document}

\title[The cosmological constant and the Higgs mass]{Seesaw relation between the cosmological constant and the Higgs mass}


\author{\fnm{R.J.} \sur{Cossins}}\email{rjcossins@protonmail.com}


\abstract{We propose the relation $M^*_{Higgs} = ({M_{\Lambda} \ M_{I}})^{\frac{1}{2}}$ where $ M^*_{Higgs}, M_{\Lambda}$ and $M_{I}$ denote the mass scale associated with the Higgs boson, the cosmological constant and the inflaton respectively. We demonstrate how this seesaw-like (geometric mean) relation perfectly matches observations and the unified scenario of holographic constant roll inflation.}




\maketitle

\section{Introduction}\label{sec1}

An effective field theory is \textit{natural} if all of its parameters are of order unity in units of the cutoff, without fine-tuning. While there may be a degree of judgement in what constitutes fine-tuning, it has long been observed that the two most significant naturalness problems are the cosmological constant problem (which is the disagreement by 120 orders of magnitude between the observed and QED field value of the vacuum energy density e.g. Ref \cite{rugh2002quantum}), and the Standard Model hierarchy problem (which is a statement about the naturalness of the mass of the Higgs boson). Notwithstanding the long and inglorious history of Higgs mass predictions (e.g. Ref \cite{Higgs-mass}), there remains a general observation that there may be a connection between the Higgs field value (and therefore the mass of the Higgs boson) and the vacuum energy density, and that a solution to both may be related to so-called \textit{UV-IR mixing}, i.e. that high and low energy physics does not decouple as expected. 

\

In regard to the cosmological constant, Ref \cite{hsu2005speculative} proposed the seesaw-like relation $M_{\Lambda} \sim (M_p \ M_s)^{\frac{1}{2}}$ where $M_{\Lambda}$ is the mass scale associated with the cosmological constant (dark energy, IR-scale), $M_s$ is a \textit{cosmic mass-energy} (closely related to the de Sitter temperature), and $M_p$ is the Planck mass-energy (UV-scale). Ref \cite{hsu2005speculative} also noted that their relation was identical to the CKN bound, which inspired the original holographic dark energy model, Ref \cite{li2004model}. In the context of de Sitter spacetime and string theory arguments, Ref \cite{berglund2023sitter, berglund2023string} recently supported Ref's \cite{hsu2005speculative} relation, and also connected the the Higgs boson mass to the cosmological constant via the seesaw-like relation $M_{Higgs}^2 \sim M_{\Lambda}  M_p $ i.e. the matter (and dark matter) scale related to the dark-energy IR and the Planck scale UV. In our Results section, we will demonstrate that this Higgs relation, while of the right form, is simplistic as it assumes the Planck-energy as the energy scale of inflation.      

\

While the details of past inflation are considered largely unknown, during inflation, all the energy was in the inflaton, which has very few degrees of freedom and low entropy, Ref \cite{egan2010larger}. Inflation is de-Sitter-like, as during inflation the proper distance $l_\Lambda$ to the cosmic event horizon (CEH) stays almost constant, while the proper distance \emph{between two points} increases exponentially. Following the end of inflation, reheating (aka the \emph{Hot Big Bang}), produces a radiation-dominated Universe. The `re’ in reheating comes from the assumption the pre-inflationary Universe was also radiation dominated. This means that we might consider the Universe before inflation as the actual initial past boundary condition, with the dS-like start of inflation being a later phase
change.

\begin{figure}[ht]
\centering
\includegraphics[scale=0.5]{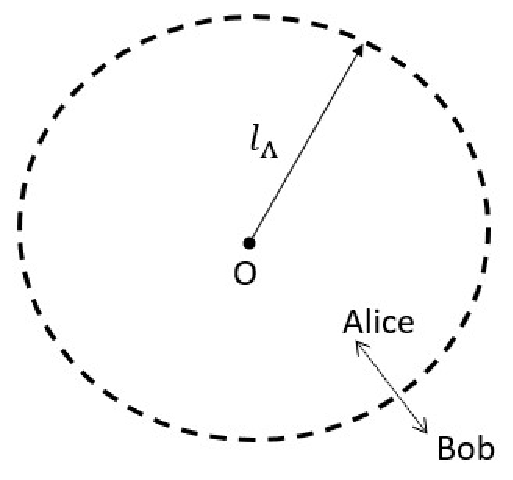}
\caption{For an observer at O inside the cosmic event horizon (CEH) with radius $l_{\Lambda}$, the universe can be divided into two sub-vacuums, $(A)$ inside the CEH and $(B)$ outside. The horizon surface $\Sigma$ has entanglement entropy $S_{dS}$ and rest energy $E_H$}.    
\label{fig:figure1}
\end{figure}

The concordance model of cosmology Lambda cold dark matter ($\Lambda$CDM), is thought to have a dS attractor and thus CEH as a far future boundary condition. Our present universe can be considered a quasi-de Sitter state, and also has a CEH due to its vacuum energy density (cosmological constant).

\section{Results}\label{sec2}
The CEH of Figure \ref{fig:figure1} is also the source of de Sitter radiation (low-energy `dark photons'), which has a specific universal minimum temperature\footnote{The factor of two difference with $T_{dS}$ the de Sitter (Unruh) and Hawking-Bekenstein temperature (and acceleration) is due to the location of the evaluation of the temperature, locally (Unruh), or remotely-at-infinity (Hawking), Ref \cite{easson2010entropic}}  and associated mass-energy via $E_{dS}=T_{dS} \ k_B$. We can relate the de Sitter energy to a sort of string energy via a basic application of string theory to Figure \ref{fig:figure1}, which gives us $E_{s}$ the string energy (dimensions $ML^{2}T^{-2}$). Here $\alpha \prime$ is the Regge slope parameter, with dimensions set as the inverse of energy squared.

\begin{equation}
\label{eq:79}
E_{s}=\frac{1}{\sqrt {\alpha \prime}} = \frac{\hslash c} {l_{\Lambda}} = m_{s}c^2
\end{equation}

The string tension $T$ (which we could also conceptualise as a nonphysical point-particle with the de Sitter mass-energy $E_{dS}/c^2 = m_{dS}=m_s/2\pi$ on a massless string), is then:

\begin{equation} 
\label{eq:2}
T= \frac{1}{2\pi \hslash c \alpha \prime} 
\end{equation}

The boundary conditions of $l_{\Lambda}$ are as a de Sitter characteristic length  defined as the past or future cosmic event horizon radius, which is also the Hubble (and Schwarzschild) radius of Figure \ref{fig:figure1}, being $l_{\Lambda} =\sqrt{ {3}/{\Lambda}}=c/H$, as per Ref \cite{bousso2002adventures}. $H$ is the future Hubble Constant (dimension T$^{-1}$), with $H_I$ being the past Hubble Constant, prior to inflation. $\Lambda$ is the cosmological constant (dimension L$^{-2}$). The string length (dimension $L$) is also:
\begin{equation}
\label{eq:81}
\hslash c \sqrt {\alpha \prime} = l_{\Lambda} = l_{s}
\end{equation}

Equivalently, a comoving volume partition of the Universe can be treated as a closed system for which $dS \geqslant0$, Ref \cite{egan2010larger}. The maximum entropy of a closed system (i.e. Figure \ref{fig:figure1}, with $L=2 \pi l_{\Lambda}$ being the circumference of the CEH), is obtained when all the energy $E_{H}$ within the system is worked, i.e. degraded into the smallest bits possible. All energy is converted into minimal energy `dark photons' with wavelengths as large as the system $L$. From the Compton wavelength relation, we can easily see that the minimum quanta of energy $E_{s}$ of this system is identical to Eq (\ref{eq:79}). Then $E_s$ is a sort of Compton rest mass-energy of the universe, Ref \cite{hsu2005speculative}.    

 \begin{equation}
\notag 
E_{s}=\frac{hc}{\lambda_{max}}=\frac{\hslash c}{l_{\Lambda}}
\end{equation}

\

The energy scale $M_{\Lambda}$ (an interaction mass-energy) can be defined from the local vacuum energy density $\rho_\Lambda = \Lambda c^2/8 \pi G$. With $\Lambda=3 \left( H_{0}/{c}\right)^2 \Omega_{\Lambda,0}$ where $\Omega_{\Lambda,0}$ is the present dark energy density parameter (dimensionless) and $H_{0}$ is the present local Hubble Constant. With the observed local $H_0 = 73.30 \pm 1.04 \ km \ s^{-1} Mpc^{-1} $ (i.e. $H_0 \approx 2.37 \times 10^{-18} \ s^{-1}$)   from Ref \cite{riess2022comprehensive} we get:

\begin{equation}
\label{eq:36}
M_{\Lambda} = \rho_\Lambda^{\frac{1}{4}} \approx 2.345\times 10^{-12} \ GeV 
\end{equation} 

We also obtain $l_{\Lambda} \approx 1.51 \times 10^{26}$m (i.e. future CEH radius). Translating the local energy scale $E_s$ from Eq (\ref{eq:79}) to a remotely-at-infinity evaluation (i.e. $m_s/2$) gives us the universal remotely-at-infinity ground-state energy scale $M_s$:

\begin{equation}
\label{eq:7}
M_{s} = \rho_s^{\frac{1}{4}} \approx 6.52\times 10^{-43} \ GeV 
\end{equation}

The Planck energy $M_p = \rho_p^{\frac{1}{4}}= 1.221\times 10^{19} \ GeV $ so, as a minor new result, we can see that the seesaw-like (i.e. geometric mean) relation from Ref \cite{hsu2005speculative} is actually:

\begin{equation}
\label{eq:38}
M_{\Lambda} = \frac{H}{H_0} \sqrt{M_p \ M_s} 
\end{equation}

This concludes our consideration of the future de-Sitter-like boundary condition. Regarding our past de-Sitter-like boundary condition (inflation), firstly, Ref's \cite{barrow2014maximum} showed that in classical General Relativity in the presence of a positive cosmological constant $\Lambda$, there are lower and upper bounds for the mass of any body:  

\begin{equation}
\label{eq:4}
m_{lower}\geqslant \frac{\hslash} {c} \sqrt{\frac{\Lambda} {3}} 
\end{equation}

\begin{equation}
\label{eq:400}
m_{upper} \leqslant \frac{c^{2}}{G}\sqrt{\frac{3}{\Lambda }}  \leqslant  \frac{c^2 l_{\Lambda}}{G}
\end{equation}

Clearly, Eq (\ref{eq:79}) and (\ref{eq:4}) are equivalent: $m_{s}=m_{lower}$. Relevant to considering the past boundary condition, Ref \cite{boehmer2005does} proved Eq (\ref{eq:4}) is equivalent to $m_{lower} \geqslant c^2 l_{\Lambda}/4G$\footnote{Do not use the future CEH length in this situation, here $l_{\Lambda}$ is the past CEH length}. Therefore, for the past boundary condition: $ m_{upper} = 2\ m_{CEH} = 4 \ m_{lower}$. Here, $m_{CEH}$ is the rest-mass energy of the cosmic event horizon in Figure \ref{fig:figure1}. For the past boundary condition we might set $m_{CEH}=m_{Planck}$. If the CEH was maximally rotating (a Kerr solution) the inner and outer CEH horizons would merge and be both equal to the Planck Length. Even so, observations do not indicate that the Universe is rotating. However, the classical length scale (Schwarzschild radius) of the quantum past ground-state energy $m_{lower}$ is the Planck Length. Therefore, at the past boundary condition $l_{\Lambda}=2L_{Planck}$.   

\

The hierarchy problem with regard to the measured/observed Higgs boson mass $m^*_{Higgs}$ can be simply expressed as an effective mass relation, where $m_{0 \ Higgs}$ is the `bare' Higgs boson mass and $S$ is the `self energy' or energy due to interactions. 
\begin{equation}
\label{eq:35}
 m^*_{Higgs} = m_{0 \ Higgs} + \frac{S}{c^2} 
\end{equation} 

The Higgs field is quantum, so one might think of the Higgs boson being the ground state quanta of this field. Further, the Higgs field has an average non-zero value, for which the energy density of the vacuum is at a minimum. The mass of the Higgs boson is determined by how quickly the energy density of the vacuum changes as you vary the Higgs field’s value away from the ground state.  If we consider the suggested relation $M_{Higgs} \sim (M_{\Lambda} \ M_p )^{\frac{1}{2}}$ from Ref \cite{berglund2023sitter}, which was presented as an upper-bound relation, we could also interpret this as stating that the effective Higgs mass is is a geometric mean of the cosmological constant and inflation-scale mass-energies. Then, defining $M_I$ as the mass-energy of an inflaton gives our main result:

\begin{equation}
\label{eq:37}
 M^*_{Higgs} = \sqrt{M_{\Lambda} \ M_{I}}
\end{equation} 

Particle Data Group \cite{PhysRevD.110.030001} gives average $m_{Higgs} \sim 125.20 \pm 0.11  \ GeV$. With this and Eq (\ref{eq:36}) we can see that $M_I \approx 6.69 \times 10^{15} \ GeV$.

\

Even without this result, it is well known that $M_p = T_{p} = 1.22 \times 10^{19} \ GeV$ cannot be the inflaton energy $M_I$. In the simplest model of inflation, the energy density at reheating equals the energy density before inflation. So, if our past boundary condition horizon length scale $l_{\Lambda}=2L_{Planck}$ sets an energy density before inflation, we might then consider that that the string Hagedorn temperature associated with this past boundary horizon length scale could be the reheating temperature. The string Hagedorn temperature $T_{H}$ can also be shown to be equivalent to the Hawking-Bekenstein black hole temperature as:
\begin{equation}
\label{eq:33}
T_{H}=\frac{1}{k_{B}4\pi \sqrt{ \alpha \prime}} =  \frac{\hslash  \kappa}{2\pi c \ k_{B}} = \frac{T_{dS}}{2} \approx 5.64\times 10^{30}K,
\end{equation}
 
Eq (\ref{eq:33}) also shows the Davis-Unruh relationship (temperature $\propto$ acceleration), via $\kappa$ the so-called black hole surface acceleration. However, $T_{H} \sim 4.86 \times 10^{17} \ GeV$ (let alone the Planck temperature of $T_{p} = 1.22 \times 10^{19} \ GeV$), is   problematic as a candidate for the reheating temperature because it is too hot. While the energy density during inflation can be greater than the reheating temperature, the reheating temperature $T_{reh}$ cannot be larger than the GUT energy scale ($\sim10^{16} \ GeV)$ otherwise relics might appear during the breaking of the GUT gauge group to the standard model gauge groups. That is, the `melting point' of quark-gluon plasma is the GUT scale. The relationship usually goes as $T_{reh} \sim V_{inf}^{1/4}$, where the potential $V_{inf} \simeq \rho_{inf}$, the energy density of inflation. Quantifying this further is model dependent, but in slow roll inflationary cosmology:

 \begin{equation}
\label{eq:700}
8 \pi G \ \rho_{inf} = 3H_I^2  \approx \frac{3}{2} \pi^2  r  A_s  M_{Pl}^4
\end{equation}

The parameter $r$ in Eq (\ref{eq:700}) is the so-called `tensor-to-scalar parameter'. Recently, $r$ has been constrained by combined observations  to $r < 0.032$ (Ref \cite{tristram2021improved}). With $\text{ln}(10^{10} A_s)\approx3.044$ from Ref \cite{akrami2018planck}, the upper bound on $\rho_{inf}$ is then:

\begin{equation}
\notag
\rho_{inf} < (1.4 \times 10^{16} \ GeV)^4
\end{equation}

\

As we have now shown what $M_I$ is not, we can now clarify what it is. We propose that our inflaton field $M_I$ must be considered as a geometric mean of $\Tilde \rho$ (the energy at the start of inflation), and $\rho_{inf}$ (the de-Sitter like energy scale during inflation). This gives our second main result:  
\begin{equation}
\label{eq:41}
 M_{I} = \sqrt  { \Tilde{\rho}^{\frac{1}{4}} \ \rho_{inf}^{\frac{1}{4}}}
\end{equation} 

We will leave further consideration of $\rho_{inf}$ aside for now, and quantify $\Tilde \rho$ in the following way. First, our Universe is not and never was a black hole, Ref \cite{lineweaver2023all}. However, the CEH means it shares many similarities. Therefore, as per the AdS/CFT duality, we consider that Universe before inflation can actually be viewed in two mathematically equivalent ways, one with gravity, one without. The Universe before inflation is a strongly gravitating system, but, is mathematically equivalent to a non-gravitational but strongly quantum system. That is, a thermal state of quantum fields, essentially, a hot plasma. This should be recognisable, it is the old assumption that the Universe was also radiation dominated before inflation, as per our Introduction.  

Then, as Ref \cite{ringwald2020gravitational} points out, there is a longstanding argument that the maximum temperature of thermal radiation is the reduced Planck mass $M_{Pl} = \sqrt{\hbar c / 8 \pi G}$ = $2.44 \times 10^{18}$ \ GeV. This then is also the lower bound temperature at which it is expected gravitons would decouple. At higher temperatures $M_{QG}>M_{Pl}$ quantum gravity becomes very important. So we might say that there is a Hagerdorn quantum gravity temperature associated with $M_{Pl}$. Inspired by Ref \cite{antoniadis2015effective}, but what is a new result, we quantify this quantum gravity energy scale $M_{QG}$ as:

\begin{equation}
\label{eq:50}
M_{Pl} = \frac {M_{QG}} {\sqrt{N_{QG}}}
\end{equation} 

We associate $N_{QG}$ with the effective degrees of freedom. As Ref \cite{carlip2017dimension} details, there are multiple lines of evidence that space-time becomes effectively 2D at distances near the Planck scale. There is also the 1988 speculation from Atick and Witten regarding a dramatic loss of degrees of freedom at the Hagerdorn phase. Also, as Ref \cite{egan2010larger} points out, entropy increased when gravitons ($g_{grav}=2$) were first produced. If we take $N_{QG}=2$ then $M_{QG} = \rho_{QG}^{\frac {1}{4}} = 3.44 \times 10^{18} \ GeV$. The associated horizon temperature is $T_{QG} \approx 5.19 \times 10^{30} K$ aka $4.47 \times 10^{17} \ GeV$, and the entropy is $S_{QG}=6\pi^2$. 

\

However, $T_{QG}$ from Eq (\ref{eq:50}) is a `bare' local temperature, so to get our desired effective remotely-at-infinity energy scale at the start of inflation $\Tilde \rho$ we first need to account for locality\footnote{As per Eq (\ref{eq:33})} (i.e. $T_{QG}/2$) and interactions. For interactions, event horizons can be modelled by a considering them as a mirror; e.g. Ref \cite{hayden2007black}. However, Ref \cite{adami2015black} has conjectured that the key difference to a mirror is that as well as reflection, event horizons also stimulate the emission of radiation in response to the incoming states that are reflected. And these stimulated states are the clones of the incoming states.

\begin{figure}[ht]
\centering
\includegraphics[scale=0.5]{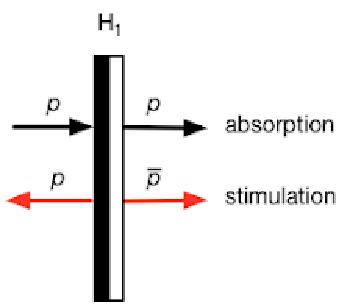}
\caption{The horizon of a perfectly absorbing black hole looks black from the outside, and white from the inside. Particle p (black) is absorbed, and stimulates the emission of the clone-anticlone pair (red) i.e. interactions of the bare BH mass with its `fluid' medium.}
\label{fig:figure7}
\end{figure}

To Alice (Figure \ref{fig:figure1}), the future CEH is a `white' horizon (Figure \ref{fig:figure7}), as only dS radiation comes out, nothing goes in. Now, black hole event horizons (and therefore, also the Figure \ref{fig:figure1} cosmic event horizon), are in fact optimal universal quantum cloners, Ref \cite{adami2015black}. By `optimal’, we mean `nearly perfect’ since perfect quantum cloning is not allowed. So from Alice's perspective, the quantum state clone fidelity is $-5/6$ (the negative
sign simply means the vector points inwards). Further, there will be an anti-clone with a fidelity of $2/3$ outside the horizon. 

From this, we obtain a heuristic that there is a local `effective velocity vector' associated with the net cloning fidelity and degrees of freedom of an event horizon as per Figure \ref{fig:figure7}, i.e. the quantum interactions. 

For a positive mass de Sitter space with a stable event horizon, accounting for the interactions gives $(2/3 - 5/6)c.$ The result is an `effective acceleration' i.e. $-cH/6$ in dS space. The negative sign simply means the vector points inwards. Or, more generally at any time $t$\footnote{With the observed local $H_0$ Eq (\ref{eq:511}) recovers the Milgrom (Ref \cite{milgrom2020a0}) acceleration scale $a_{ef,0}=a_{m} \approx 1.2\times 10^{-10} m \ s^{-2}$}:  

\begin{equation}
\label{eq:511}
a_{ef,t} = \frac{a}{6\sqrt{\Omega_{\Lambda,t}}}=\frac{cH_{t}}{6}
\end{equation}

With $T_{QG}/2$ (locality) and Eq (\ref{eq:511}) (interactions), we get the following original result:

\begin{equation}
\label{eq:51}
\Tilde \rho^{\frac{1}{4}} = \frac{T_{QG}}{2 \times 6 \sqrt{\Omega_{\Lambda,I}}}
\end{equation} 

As during inflation we can take $\Omega_{\Lambda,I} = 1$ numerically Eq (\ref{eq:51}) gives $\Tilde \rho^{\frac{1}{4}} = 3.73 \times 10^{16} \ GeV$. Plugging this result into Eq (\ref{eq:41}) with Eq (\ref{eq:37}), we obtain $\rho_{inf}^{\frac{1}{4}} \approx 1.2 \times 10^{15} \ GeV$.   

 \section{Discussion}\label{sec3}
While the above results may be interesting to some, they might also appear to many as overly speculative or simply numerical coincidences (e.g. as per Ref \cite{bogan2009cosmological}), lacking a unified mathematical model. However, a modified holographic dark energy model provides does in fact provide a unifying framework for our results. Now, the original holographic dark energy (HDE) cosmological model by Ref \cite{li2004model} was motivated as a solution to the cosmological constant problem. 
\begin{equation}
\label{eq:70}
\rho _{\Lambda} = \frac{3c^2}{8\pi G L^2}. C^2  
\end{equation} 
Here $c$ is the (dimensionless) model parameter of the HDE which is positive and (generally) taken as constant and $C$ is the speed of light. As the Friedmann equation is $\rho_{\Lambda}=3M_p^2H^2$ you get (in natural units) $c=H \ L$ with $L$ as the future Hubble horizon which equals the future event horizon, which is why Ref \cite{li2004model} felt the HDE parameter $c=1$. The HDE model is defined in terms of an IR cut-off $L$. The future event horizon is the common choice for $L$. This gives an EoS:

\begin{equation}
\label{eq:71}
  w_{\Lambda}=-\frac{1}{3}\left( 1+\frac{2\sqrt{\Omega _{\Lambda}}}{c}\right)  
\end{equation} 

And this is where the problems begin (e.g. Ref \cite{colgain2021critique}), because with Planck 2018 \cite{aghanim2021planck} $\Omega_{\Lambda,0} = 0.6889 \pm 0.0056$ Eq (\ref{eq:71}) gives $w_0=-0.89$ which does not match Planck 2018 \cite{aghanim2021planck} $w_0= -1.03 \pm 0.03$. However, if we consider that $c$ varies with time, such that the Hubble horizon is the IR cut-off $L$, we still get an accelerating Universe, as per Ref \cite{malekjani2018can}. However, unlike Ref \cite{malekjani2018can}, we propose a simple time-varying model i.e. $c(t)=H/H_0=\sqrt\Omega_{\Lambda}$. This gives $w = -1$ with $c_\infty=1$. As this version of HDE mirrors $\Lambda$CDM, it must also be considered to be similar to a HDE model with a constant, fine-tuned parameter $c$. 

\

\begin{figure}[ht]
\centering
\includegraphics[scale=0.5]{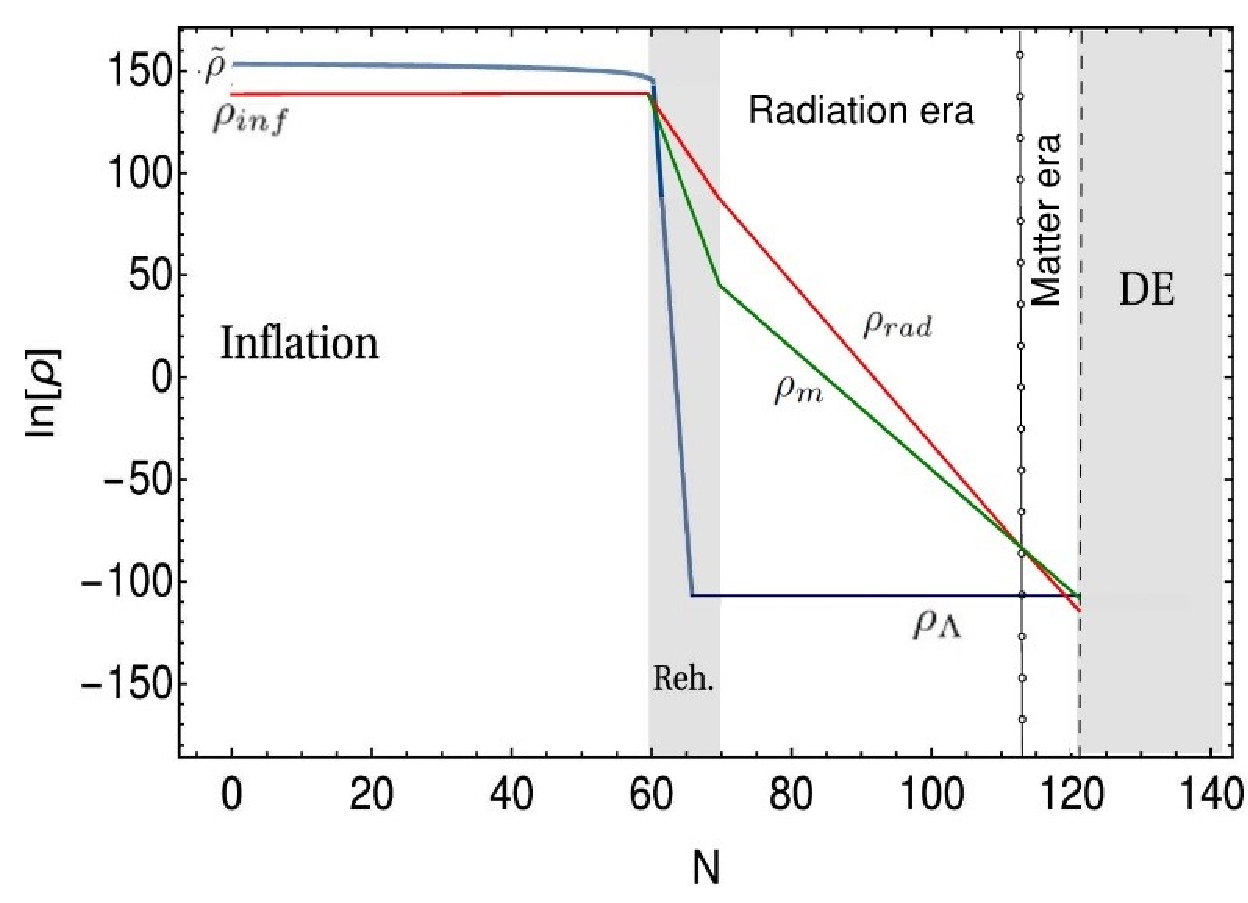}
\caption{$\ln \Tilde{\rho}$ (the blue curve), $\ln \rho_{inf}$ (the red curve) vs. the $e$-fold variable N from the beginning of inflation to the current day (vertical dashed line), adapting the unified HDE model of Ref \cite{nojiri2023holographic}. The (dark+baryonic) matter density $\ln\rho_m$ (the green curve) is interpolated back to the end of reheating via $N_0$ obtained from $\rho_{rad,0}$ and $\rho_m(N)=\rho_m^{(re)}e^{-3[N-(N_f+N_{re})]}$}    
\label{fig:figure3}
\end{figure}

Recently, Ref \cite{nojiri2023holographic} proposed a unified holographic scenario, a generalised form of HDE that gives all known forms of HDE as particular cases combined with constant roll inflation. We would expect generalised HDE to also include our simple HDE proposal above as a particular case. In fact, the unified holographic scenario of Ref \cite{nojiri2023holographic} does produce extremely similar results to what we have derived, even though their model does not quantify a decay channel to  produce matter as well as radiation (we have only roughly estimated the matter curve in Figure \ref{fig:figure3}, further work is needed). The key point, as shown in  Figure \ref{fig:figure3}, is that in unified holographic constant roll inflation there are \textit{two} inflation-related energy scales; $\Tilde \rho$ and $\rho_{inf}$\footnote{$\rho_{inf}$ = $\rho_n$ in Ref \cite{nojiri2023holographic}}.

The energy scale $\Tilde \rho$ is initially associated with the cut-off $\Tilde L_{IR}$, the Hubble radius at the start of inflation. This energy scale decays exponentially during inflation (possibly tracking the time it takes for quantum complexity equilibrium of the interior to be reached). $\Tilde \rho$ then decays even more rapidly (but not instantaneously), during post-inflation reheating, finally reaching a base level near the end of reheating, i.e. $ \Tilde \rho = \rho_{\Lambda}$. Then $\rho_{\Lambda}$ remains almost constant over the radiation, matter and dark energy cosmological eras. That is, late-time acceleration (our current era), is controlled by an almost constant dark-energy density (likely undergoing quantum complexification), associated now with our future Hubble radius.  
 
The energy scale $\rho_{inf}$ is initially associated with $ L_{n}$, another Hubble radius throughout inflation. This is the `inflation dark-energy' density which \textit{does} stay almost constant during inflation, but has a coupling with radiation (and matter), decaying only once reheating starts (thermalisation phase, with energy being transferred into a relativistic fluid), with the reheating temperature providing the initial\footnote{A possible natural solution to providing some \textit{early dark radiation}} energy density in the radiation dominated era: $p_{inf} \rightarrow \rho_{rad}$. The later inflection point where $\rho_{rad} = \rho_{m}$ marks the start of the relatively short matter era, followed by dark energy domination and the present day. In fact, $\rho_{inf}$ is still decaying at the same rate today, (i.e. thermalisation, marked by the increasing entanglement entropy of the CEH). 

\

The unified holographic scenario is consistent with observational constraints by selecting suitable estimated ranges of model parameters.  In particular, Ref \cite{nojiri2023holographic} required $\Tilde \rho \sim 10^{66} \ GeV^4$ and $\rho_{inf} \sim 10^{61} \ GeV^4$ as initial conditions $N=0$. This matches very well with our results of $\Tilde \rho^{\frac{1}{4}} = 3.73 \times 10^{16} \ GeV$ (aka $\Tilde \rho \sim 1.93 \times 10^{66} \ GeV^4$) and $\rho_{inf}^{\frac{1}{4}} \approx 1.2 \times 10^{15} \ GeV$ (aka $\rho_{inf} \sim 2.06 \times 10^{60} \ GeV^4$). 
 
\ 

With this, we can actually wring some more from our Results section. Firstly, the entropy $S_{QG}$ also sets a precise bound on constant-roll inflation. That is, $\in_1=1$ when the the number of $e$-foldings during inflation $N_{inf}=6\pi^2 \approx 59$. It is commonly recognised that about $60$ $e$-folds are required to solve the horizon, flatness and monopole problems in cosmology, which is what Ref \cite{nojiri2023holographic} used.  Further support for our Results can also be found via Ref \cite{padmanabhan2016we} identical reported estimate of the number of $e$-foldings during inflation. Ref \cite{Padmanabhan_2014, padmanabhan2016we} also demonstrated how the classical Friedmann equation can be written in terms of three key energy densities, $\rho_{inf}$ density during inflation, $\rho_{eq}$ density at matter-radiation equilibrium and $\rho_\Lambda$ dark energy density at late times. 
Also, as per Ref \cite{phillips2014holographic}, you can make a holographic-inspired argument that there is a limit to the information $I_c$ available to an eternal observer, which also places a bound on the amount of inflation. This informational limit is represented by the quantum state before inflation, and the result is $4\pi$ nats by information theoretic arguments similar to those used to derive the entropy of a black hole. Ref \cite{Padmanabhan_2014, padmanabhan2016we} also presented an interesting holographic-inspired conjecture relating $I_c$ to the three key energy scales $\rho_{inf}$, $\rho_{eq}$ and $\rho_\Lambda$:  
\begin{equation}
\label{eq:55}
I_c = \frac{1}{9 \pi} \ln \left (\frac{4}{27} \frac{\rho_{inf}^{3/2}}{\rho_{\Lambda} \ \rho_{eq}^{1/2}}  \right)
\end{equation} 

Ref \cite{Padmanabhan_2014, padmanabhan2016we} supported this conjecture with observational results for $\rho_{\Lambda}$ and $\rho_{eq}$ plus an informed estimate of $\rho_{inf}^{\frac{1}{4}}\approx 1.2 \times 10^{15} \ GeV$ (very similar to our Results), showing $I_c\approx4\pi$.  

\

After the end of inflation, if the Universe enters a Kamionkowski-like reheating phase, reheating is not instantaneous, and will be characterised  by a reheating $e$-fold number $N_{re}$. We adapt a holographic bound argument from Ref \cite{phillips2014holographic} regarding the number of $e$-folds of inflation, such that $N_t=N_{inf}+N_{re}$.

\begin{equation}
\label{eq:115}
N_t = \frac{1}{3} \left( \ 3 - \frac{2}{1+w} \right) ln\frac{H_{QG}} {H}
\end{equation} 

Here, $H$ is the future Hubble constant, and $H_{QG}$ is the Hubble constant at the start of inflation, and in the case of radiation $w=1/3$. Therefore, $N_t \approx 69.46$ giving $N_{re} \approx 10.2$. In Ref \cite{nojiri2023holographic} $N_{inf} = 60$ and $N_{re} \approx 10$. 

\

Also, the Planck energy scale $M_{p} = 1.22 \times 10^{19} \ GeV$ is related to Eq (\ref{eq:50}) via:

\begin{equation}
\label{eq:57}
M_{QG} = \frac {M_{p}} {\sqrt{N_{p}}}
\end{equation} 

From Eq (\ref{eq:57}) we see that the effective degrees of freedom $N_{p}=4S_{p}=S_{I}=4 \pi$, the surface area of a sphere with $r=L_{Planck}$, an instanton. The holographic principle equates the actual observed degrees of freedom of a bulk to its area not its volume. The energy density $M_r  = 7.17 \times 10^{18} \ GeV$  of the instanton is also related to the energy density $M_{l_{\Lambda}} = 5.07 \times 10^{18} \ GeV$ of our length scale $l_{\Lambda}$, implying space-time is 2D from $M_r$: 

\begin{equation}
\label{eq:58}
M_{l_{\Lambda}} = \frac {M_{r}} {\sqrt{N_{QG}}}
\end{equation} 

So we might describe, more precisely, the past boundary condition from our Introduction section equivalently as: the entropy $S_I$ of the CEH just before inflation, the effective degrees of freedom of an instanton $N_p$ (possibly the initial condition of our Universe), a proportionality constant in natural units: the CEH circumference divided by the CEH rest energy, or, the information available to an eternal observer $I_c$.

 \section{Conclusions}\label{sec8}

Of course, the approaches we use in this work are all semi-classical, so our results still only approximates QFT quantum corrections to the self energy terms - we do not have a `final theory' here. While we hope that this paper is considered a sufficiently intriguing original contribution, our results are more of a suggestion for future work. Not least, that while the unified scenario implies dark energy density is almost constant, defining \textit{almost} (e.g. Ref \cite{adame2024desi}) remains an enormous problem. While we used precision cosmological observations plus particle physics experimental results to support our main result $ M^*_{Higgs} = \sqrt{M_{\Lambda} \ M_{I}}$, the obvious question is whether we could successfully estimate $ M^*_{Higgs}$ solely from minimal cosmological observations as free parameters. Planned work on the unified HDE scenario, plus Eq (\ref{eq:55}) and Eq (\ref{eq:511}) could provide some angles. However, one would also need to successfully attack the Hubble Tension problem, unresolved despite a multitude of attempts (e.g. Ref \cite{di2021realm}). This, we leave for future work.

\section*{Data availability statement}
All data generated or analysed during this study are included in this published article

\section*{Acknowledgments}

The author thanks private conversations which helped us improve upon an earlier draft of this paper. No funding was received for conducting this study.





\bibliography{sn-bibliography}


\begin{thebibliography}{32}
\ifx \bisbn   \undefined \def \bisbn  #1{ISBN #1}\fi
\ifx \binits  \undefined \def \binits#1{#1}\fi
\ifx \bauthor  \undefined \def \bauthor#1{#1}\fi
\ifx \batitle  \undefined \def \batitle#1{#1}\fi
\ifx \bjtitle  \undefined \def \bjtitle#1{#1}\fi
\ifx \bvolume  \undefined \def \bvolume#1{\textbf{#1}}\fi
\ifx \byear  \undefined \def \byear#1{#1}\fi
\ifx \bissue  \undefined \def \bissue#1{#1}\fi
\ifx \bfpage  \undefined \def \bfpage#1{#1}\fi
\ifx \blpage  \undefined \def \blpage #1{#1}\fi
\ifx \burl  \undefined \def \burl#1{\textsf{#1}}\fi
\ifx \doiurl  \undefined \def \doiurl#1{\url{https://doi.org/#1}}\fi
\ifx \betal  \undefined \def \betal{\textit{et al.}}\fi
\ifx \binstitute  \undefined \def \binstitute#1{#1}\fi
\ifx \binstitutionaled  \undefined \def \binstitutionaled#1{#1}\fi
\ifx \bctitle  \undefined \def \bctitle#1{#1}\fi
\ifx \beditor  \undefined \def \beditor#1{#1}\fi
\ifx \bpublisher  \undefined \def \bpublisher#1{#1}\fi
\ifx \bbtitle  \undefined \def \bbtitle#1{#1}\fi
\ifx \bedition  \undefined \def \bedition#1{#1}\fi
\ifx \bseriesno  \undefined \def \bseriesno#1{#1}\fi
\ifx \blocation  \undefined \def \blocation#1{#1}\fi
\ifx \bsertitle  \undefined \def \bsertitle#1{#1}\fi
\ifx \bsnm \undefined \def \bsnm#1{#1}\fi
\ifx \bsuffix \undefined \def \bsuffix#1{#1}\fi
\ifx \bparticle \undefined \def \bparticle#1{#1}\fi
\ifx \barticle \undefined \def \barticle#1{#1}\fi
\bibcommenthead
\ifx \bconfdate \undefined \def \bconfdate #1{#1}\fi
\ifx \botherref \undefined \def \botherref #1{#1}\fi
\ifx \url \undefined \def \url#1{\textsf{#1}}\fi
\ifx \bchapter \undefined \def \bchapter#1{#1}\fi
\ifx \bbook \undefined \def \bbook#1{#1}\fi
\ifx \bcomment \undefined \def \bcomment#1{#1}\fi
\ifx \oauthor \undefined \def \oauthor#1{#1}\fi
\ifx \citeauthoryear \undefined \def \citeauthoryear#1{#1}\fi
\ifx \endbibitem  \undefined \def \endbibitem {}\fi
\ifx \bconflocation  \undefined \def \bconflocation#1{#1}\fi
\ifx \arxivurl  \undefined \def \arxivurl#1{\textsf{#1}}\fi
\csname PreBibitemsHook\endcsname

\bibitem{rugh2002quantum}
\begin{barticle}
\bauthor{\bsnm{Rugh}, \binits{S.E.}},
\bauthor{\bsnm{Zinkernagel}, \binits{H.}}:
\batitle{The quantum vacuum and the cosmological constant problem}.
\bjtitle{Studies In History and Philosophy of Science Part B: Studies In History and Philosophy of Modern Physics}
\bvolume{33}(\bissue{4}),
\bfpage{663}--\blpage{705}
(\byear{2002}).
\doiurl{10.48550/arXiv.hep-th/0012253}
\end{barticle}
\endbibitem

\bibitem{Higgs-mass}
\begin{barticle}
\bauthor{\bsnm{Schucker}, \binits{T.}}:
\batitle{Higgs-mass predictions}.
\bjtitle{arXiv preprint arXiv:0708.3344}
(\byear{2011}).
\doiurl{10.48550/arXiv.0708.3344}
\end{barticle}
\endbibitem

\bibitem{hsu2005speculative}
\begin{barticle}
\bauthor{\bsnm{Hsu}, \binits{S.}},
\bauthor{\bsnm{Zee}, \binits{A.}}:
\batitle{A speculative relation between the cosmological constant and the planck mass}.
\bjtitle{Modern Physics Letters A}
\bvolume{20}(\bissue{35}),
\bfpage{2699}--\blpage{2703}
(\byear{2005}).
\doiurl{10.1142/S0217732305018839}
\end{barticle}
\endbibitem

\bibitem{li2004model}
\begin{barticle}
\bauthor{\bsnm{Li}, \binits{M.}}:
\batitle{A model of holographic dark energy}.
\bjtitle{Physics Letters B}
\bvolume{603}(\bissue{1-2}),
\bfpage{1}--\blpage{5}
(\byear{2004}).
\doiurl{10.48550/arXiv.hep-th/0403127}
\end{barticle}
\endbibitem

\bibitem{berglund2023sitter}
\begin{barticle}
\bauthor{\bsnm{Berglund}, \binits{P.}},
\bauthor{\bsnm{H{\"u}bsch}, \binits{T.}},
\bauthor{\bsnm{Minic}, \binits{D.}}:
\batitle{On de sitter spacetime and string theory}.
\bjtitle{International Journal of Modern Physics D}
\bvolume{32}(\bissue{09}),
\bfpage{2330002}
(\byear{2023}).
\doiurl{10.48550/arXiv.2212.06086}
\end{barticle}
\endbibitem

\bibitem{berglund2023string}
\begin{barticle}
\bauthor{\bsnm{Berglund}, \binits{P.}},
\bauthor{\bsnm{H{\"u}bsch}, \binits{T.}},
\bauthor{\bsnm{Minic}, \binits{D.}}:
\batitle{String theory bounds on the cosmological constant, the higgs mass, and the quark and lepton masses}.
\bjtitle{Symmetry}
\bvolume{15}(\bissue{9}),
\bfpage{1660}
(\byear{2023}).
\doiurl{10.3390/sym15091660}
\end{barticle}
\endbibitem

\bibitem{egan2010larger}
\begin{barticle}
\bauthor{\bsnm{Egan}, \binits{C.A.}},
\bauthor{\bsnm{Lineweaver}, \binits{C.H.}}:
\batitle{A larger estimate of the entropy of the universe}.
\bjtitle{The Astrophysical Journal}
\bvolume{710}(\bissue{2}),
\bfpage{1825}--\blpage{1834}
(\byear{2010}).
\doiurl{10.1088/0004-637x/710/2/1825}
\end{barticle}
\endbibitem

\bibitem{easson2010entropic}
\begin{barticle}
\bauthor{\bsnm{Easson}, \binits{D.A.}},
\bauthor{\bsnm{Frampton}, \binits{P.H.}},
\bauthor{\bsnm{Smoot}, \binits{G.F.}}:
\batitle{Entropic accelerating universe}.
\bjtitle{Physics Letters B}
\bvolume{696}(\bissue{3}),
\bfpage{273}--\blpage{277}
(\byear{2011}).
\doiurl{10.1016/j.physletb.2010.12.025}
\end{barticle}
\endbibitem

\bibitem{bousso2002adventures}
\begin{barticle}
\bauthor{\bsnm{Bousso}, \binits{R.}}:
\batitle{Adventures in de sitter space}.
\bjtitle{arXiv preprint hep-th/0205177}
(\byear{2002}).
\doiurl{10.48550/arXiv.hep-th/0205177}
\end{barticle}
\endbibitem

\bibitem{riess2022comprehensive}
\begin{barticle}
\bauthor{\bsnm{Riess}, \binits{A.G.}},
\bauthor{\bsnm{Yuan}, \binits{W.}},
\bauthor{\bsnm{Macri}, \binits{L.M.}},
\bauthor{\bsnm{Scolnic}, \binits{D.}},
\bauthor{\bsnm{Brout}, \binits{D.}},
\bauthor{\bsnm{Casertano}, \binits{S.}},
\bauthor{\bsnm{Jones}, \binits{D.O.}},
\bauthor{\bsnm{Murakami}, \binits{Y.}},
\bauthor{\bsnm{Anand}, \binits{G.S.}},
\bauthor{\bsnm{Breuval}, \binits{L.}}, \betal:
\batitle{A comprehensive measurement of the local value of the hubble constant with 1 km s- 1 mpc- 1 uncertainty from the hubble space telescope and the sh0es team}.
\bjtitle{The Astrophysical journal letters}
\bvolume{934}(\bissue{1}),
\bfpage{7}
(\byear{2022}).
\doiurl{10.3847/2041-8213/ac5c5b}
\end{barticle}
\endbibitem

\bibitem{barrow2014maximum}
\begin{barticle}
\bauthor{\bsnm{Barrow}, \binits{J.D.}},
\bauthor{\bsnm{Gibbons}, \binits{G.W.}}:
\batitle{Maximum tension: with and without a cosmological constant}.
\bjtitle{Monthly Notices of the Royal Astronomical Society}
\bvolume{446}(\bissue{4}),
\bfpage{3874}--\blpage{3877}
(\byear{2014}).
\doiurl{10.1093/mnras/stu2378}
\end{barticle}
\endbibitem

\bibitem{boehmer2005does}
\begin{barticle}
\bauthor{\bsnm{Böhmer}, \binits{C.G.}},
\bauthor{\bsnm{Harko}, \binits{T.}}:
\batitle{Does the cosmological constant imply the existence of a minimum mass?}
\bjtitle{Physics Letters B}
\bvolume{630}(\bissue{3–4}),
\bfpage{73}--\blpage{77}
(\byear{2005}).
\doiurl{10.1016/j.physletb.2005.09.071}
\end{barticle}
\endbibitem

\bibitem{PhysRevD.110.030001}
\begin{barticle}
\bauthor{\bsnm{Navas}, \binits{S.}}, \betal:
\batitle{Review of particle physics}.
\bjtitle{Phys. Rev. D}
\bvolume{110},
\bfpage{030001}
(\byear{2024}).
\doiurl{10.1103/PhysRevD.110.030001}
\end{barticle}
\endbibitem

\bibitem{tristram2021improved}
\begin{barticle}
\bauthor{\bsnm{Tristram}, \binits{M.}},
\bauthor{\bsnm{Banday}, \binits{A.J.}},
\bauthor{\bsnm{G{\'o}rski}, \binits{K.M.}},
\bauthor{\bsnm{Keskitalo}, \binits{R.}},
\bauthor{\bsnm{Lawrence}, \binits{C.}},
\bauthor{\bsnm{Andersen}, \binits{K.J.}},
\bauthor{\bsnm{Barreiro}, \binits{R.B.}},
\bauthor{\bsnm{Borrill}, \binits{J.}},
\bauthor{\bsnm{Colombo}, \binits{L.}},
\bauthor{\bsnm{Eriksen}, \binits{H.}}, \betal:
\batitle{Improved limits on the tensor-to-scalar ratio using bicep and p lanck data}.
\bjtitle{Physical Review D}
\bvolume{105}(\bissue{8}),
\bfpage{083524}
(\byear{2022}).
\doiurl{10.1103/PhysRevD.105.083524}
\end{barticle}
\endbibitem

\bibitem{akrami2018planck}
\begin{barticle}
\bauthor{\bsnm{Akrami}, \binits{Y.}},
\bauthor{\bsnm{Arroja}, \binits{F.}},
\bauthor{\bsnm{Ashdown}, \binits{M.}},
\bauthor{\bsnm{Aumont}, \binits{J.}},
\bauthor{\bsnm{Baccigalupi}, \binits{C.}},
\bauthor{\bsnm{Ballardini}, \binits{M.}},
\bauthor{\bsnm{Banday}, \binits{A.J.}},
\bauthor{\bsnm{Barreiro}, \binits{R.}},
\bauthor{\bsnm{Bartolo}, \binits{N.}},
\bauthor{\bsnm{Basak}, \binits{S.}}, \betal:
\batitle{Planck 2018 results-x. constraints on inflation}.
\bjtitle{Astronomy \& Astrophysics}
\bvolume{641},
\bfpage{10}
(\byear{2020}).
\doiurl{10.1051/0004-6361/201833887}
\end{barticle}
\endbibitem

\bibitem{lineweaver2023all}
\begin{barticle}
\bauthor{\bsnm{Lineweaver}, \binits{C.H.}},
\bauthor{\bsnm{Patel}, \binits{V.M.}}:
\batitle{All objects and some questions}.
\bjtitle{American Journal of Physics}
\bvolume{91}(\bissue{10}),
\bfpage{819}--\blpage{825}
(\byear{2023}).
\doiurl{10.1119/5.0150209}
\end{barticle}
\endbibitem

\bibitem{ringwald2020gravitational}
\begin{barticle}
\bauthor{\bsnm{Ringwald}, \binits{A.}},
\bauthor{\bsnm{Schütte-Engel}, \binits{J.}},
\bauthor{\bsnm{Tamarit}, \binits{C.}}:
\batitle{Gravitational waves as a big bang thermometer}.
\bjtitle{Journal of Cosmology and Astroparticle Physics}
\bvolume{2021}(\bissue{03}),
\bfpage{054}
(\byear{2021}).
\doiurl{10.1088/1475-7516/2021/03/054}
\end{barticle}
\endbibitem

\bibitem{antoniadis2015effective}
\begin{barticle}
\bauthor{\bsnm{Antoniadis}, \binits{I.}},
\bauthor{\bsnm{Patil}, \binits{S.P.}}:
\batitle{The effective planck mass and the scale of inflation}.
\bjtitle{The European Physical Journal C}
\bvolume{75},
\bfpage{1}--\blpage{12}
(\byear{2015}).
\doiurl{10.1140/epjc/s10052-015-3411-z}
\end{barticle}
\endbibitem

\bibitem{carlip2017dimension}
\begin{barticle}
\bauthor{\bsnm{Carlip}, \binits{S.}}:
\batitle{Dimension and dimensional reduction in quantum gravity}.
\bjtitle{Classical and Quantum Gravity}
\bvolume{34}(\bissue{19}),
\bfpage{193001}
(\byear{2017}).
\doiurl{10.1088/1361-6382/aa8535}
\end{barticle}
\endbibitem

\bibitem{hayden2007black}
\begin{barticle}
\bauthor{\bsnm{Hayden}, \binits{P.}},
\bauthor{\bsnm{Preskill}, \binits{J.}}:
\batitle{Black holes as mirrors: quantum information in random subsystems}.
\bjtitle{Journal of High Energy Physics}
\bvolume{2007}(\bissue{09}),
\bfpage{120}--\blpage{120}
(\byear{2007}).
\doiurl{10.1088/1126-6708/2007/09/120}
\end{barticle}
\endbibitem

\bibitem{adami2015black}
\begin{barticle}
\bauthor{\bsnm{Adami}, \binits{C.}},
\bauthor{\bsnm{Steeg}, \binits{G.V.}}:
\batitle{Black holes are almost optimal quantum cloners}.
\bjtitle{Journal of Physics A: Mathematical and Theoretical}
\bvolume{48}(\bissue{23}),
\bfpage{23}--\blpage{01}
(\byear{2015}).
\doiurl{10.1088/1751-8113/48/23/23ft01}
\end{barticle}
\endbibitem

\bibitem{milgrom2020a0}
\begin{barticle}
\bauthor{\bsnm{Milgrom}, \binits{M.}}:
\batitle{The a0—cosmology connection in mond. arxiv 2020}.
\bjtitle{arXiv preprint arXiv:2001.09729}
(\byear{2020}).
\doiurl{10.48550/arXiv.2001.09729}
\end{barticle}
\endbibitem

\bibitem{bogan2009cosmological}
\begin{barticle}
\bauthor{\bsnm{Bogan}, \binits{J.R.}}:
\batitle{From the cosmological constant: Higgs boson, dark matter, and quantum gravity scales}.
\bjtitle{arXiv preprint arXiv:0902.2600}
(\byear{2009}).
\doiurl{10.48550/arXiv.0902.2600}
\end{barticle}
\endbibitem

\bibitem{colgain2021critique}
\begin{barticle}
\bauthor{\bsnm{Colg{\'a}in}, \binits{E.{\'O}.}},
\bauthor{\bsnm{Sheikh-Jabbari}, \binits{M.}}:
\batitle{A critique of holographic dark energy}.
\bjtitle{Classical and Quantum Gravity}
\bvolume{38}(\bissue{17}),
\bfpage{177001}
(\byear{2021}).
\doiurl{10.1088/1361-6382/ac1504}
\end{barticle}
\endbibitem

\bibitem{aghanim2021planck}
\begin{barticle}
\bauthor{\bsnm{Aghanim}, \binits{N.}},
\bauthor{\bsnm{Akrami}, \binits{Y.}},
\bauthor{\bsnm{Ashdown}, \binits{M.}},
\bauthor{\bsnm{Aumont}, \binits{J.}},
\bauthor{\bsnm{Baccigalupi}, \binits{C.}},
\bauthor{\bsnm{Ballardini}, \binits{M.}},
\bauthor{\bsnm{Banday}, \binits{A.J.}},
\bauthor{\bsnm{Barreiro}, \binits{R.}},
\bauthor{\bsnm{Bartolo}, \binits{N.}},
\bauthor{\bsnm{Basak}, \binits{S.}}, \betal:
\batitle{Planck 2018 results-vi. cosmological parameters}.
\bjtitle{Astronomy \& Astrophysics}
\bvolume{641},
\bfpage{6}
(\byear{2020}).
\doiurl{10.1051/0004-6361/201833910}
\end{barticle}
\endbibitem

\bibitem{malekjani2018can}
\begin{barticle}
\bauthor{\bsnm{Malekjani}, \binits{M.}},
\bauthor{\bsnm{Rezaei}, \binits{M.}},
\bauthor{\bsnm{Akhlaghi}, \binits{I.A.}}:
\batitle{Can holographic dark energy models fit the observational data?}
\bjtitle{Physical Review D}
\bvolume{98}(\bissue{6}),
\bfpage{063533}
(\byear{2018}).
\doiurl{10.1103/PhysRevD.98.063533}
\end{barticle}
\endbibitem

\bibitem{nojiri2023holographic}
\begin{barticle}
\bauthor{\bsnm{Nojiri}, \binits{S.}},
\bauthor{\bsnm{Odintsov}, \binits{S.D.}},
\bauthor{\bsnm{Paul}, \binits{T.}}:
\batitle{Holographic realization of constant roll inflation and dark energy: An unified scenario}.
\bjtitle{Physics Letters B}
\bvolume{841},
\bfpage{137926}
(\byear{2023}).
\doiurl{10.1016/j.physletb.2023.137926}
\end{barticle}
\endbibitem

\bibitem{padmanabhan2016we}
\begin{barticle}
\bauthor{\bsnm{Padmanabhan}, \binits{T.}}:
\batitle{Do we really understand the cosmos?}
\bjtitle{Comptes Rendus. Physique}
\bvolume{18}(\bissue{3–4}),
\bfpage{275}--\blpage{291}
(\byear{2017}).
\doiurl{10.1016/j.crhy.2017.02.001}
\end{barticle}
\endbibitem

\bibitem{Padmanabhan_2014}
\begin{barticle}
\bauthor{\bsnm{Padmanabhan}, \binits{T.}},
\bauthor{\bsnm{Padmanabhan}, \binits{H.}}:
\batitle{Cosmological constant from the emergent gravity perspective}.
\bjtitle{International Journal of Modern Physics D}
\bvolume{23}(\bissue{06}),
\bfpage{1430011}
(\byear{2014}).
\doiurl{10.1142/s0218271814300110}
\end{barticle}
\endbibitem

\bibitem{phillips2014holographic}
\begin{botherref}
\oauthor{\bsnm{Phillips}, \binits{D.}},
\oauthor{\bsnm{Scacco}, \binits{A.}},
\oauthor{\bsnm{Albrecht}, \binits{A.}}:
Holographic bounds and finite inflation.
Physical Review D
\textbf{91}(4)
(2015).
\doiurl{10.1103/physrevd.91.043513}
\end{botherref}
\endbibitem

\bibitem{adame2024desi}
\begin{barticle}
\bauthor{\bsnm{Adame}, \binits{A.}},
\bauthor{\bsnm{Aguilar}, \binits{J.}},
\bauthor{\bsnm{Ahlen}, \binits{S.}},
\bauthor{\bsnm{Alam}, \binits{S.}},
\bauthor{\bsnm{Alexander}, \binits{D.}},
\bauthor{\bsnm{Alvarez}, \binits{M.}},
\bauthor{\bsnm{Alves}, \binits{O.}},
\bauthor{\bsnm{Anand}, \binits{A.}},
\bauthor{\bsnm{Andrade}, \binits{U.}},
\bauthor{\bsnm{Armengaud}, \binits{E.}}, \betal:
\batitle{Desi 2024 vi: Cosmological constraints from the measurements of baryon acoustic oscillations}.
\bjtitle{arXiv preprint arXiv:2404.03002}
(\byear{2024}).
\doiurl{10.48550/arXiv.2404.03002}
\end{barticle}
\endbibitem

\bibitem{di2021realm}
\begin{barticle}
\bauthor{\bsnm{Di~Valentino}, \binits{E.}},
\bauthor{\bsnm{Mena}, \binits{O.}},
\bauthor{\bsnm{Pan}, \binits{S.}},
\bauthor{\bsnm{Visinelli}, \binits{L.}},
\bauthor{\bsnm{Yang}, \binits{W.}},
\bauthor{\bsnm{Melchiorri}, \binits{A.}},
\bauthor{\bsnm{Mota}, \binits{D.F.}},
\bauthor{\bsnm{Riess}, \binits{A.G.}},
\bauthor{\bsnm{Silk}, \binits{J.}}:
\batitle{In the realm of the hubble tension—a review of solutions}.
\bjtitle{Classical and Quantum Gravity}
\bvolume{38}(\bissue{15}),
\bfpage{153001}
(\byear{2021}).
\doiurl{10.1088/1361-6382/ac086d}
\end{barticle}
\endbibitem

\end{thebibliography}


\end{document}